\documentclass[12pt]{ucthesis}
\makeatletter
\renewcommand{\normalsize}{\@setfontsize\normalsize\@xipt{13.6}%
\abovedisplayskip 11\p@ plus3\p@ minus6\p@
\belowdisplayskip \abovedisplayskip
\abovedisplayshortskip  \z@ plus3\p@  
\belowdisplayshortskip  6.5\p@ plus3.5\p@ minus3\p@
\let\@listi\@listI}   % Setting of \@listi added 9 Jun 87
\renewcommand{\small}{\@setfontsize\small\@xpt{12}%
\abovedisplayskip 10\p@ plus2\p@ minus5\p@
\belowdisplayskip \abovedisplayskip
\abovedisplayshortskip  \z@ plus3\p@  
\belowdisplayshortskip  6\p@ plus3\p@ minus3\p@
\def\@listi{\leftmargin\leftmargini %% Added 22 Dec 87
\topsep 6\p@ plus2\p@ minus2\p@\parsep 3\p@ plus2\p@ minus\p@
\itemsep \parsep}}
\renewcommand{\footnotesize}{\@setfontsize\footnotesize\@ixpt{11}%
\abovedisplayskip 8\p@ plus2\p@ minus4\p@
\belowdisplayskip \abovedisplayskip
\abovedisplayshortskip \z@ plus\p@
\belowdisplayshortskip 4\p@ plus2\p@ minus2\p@
\def\@listi{\leftmargin\leftmargini %% Added 22 Dec 87
\topsep 4\p@ plus2\p@ minus2\p@\parsep 2\p@ plus\p@ minus\p@
\itemsep \parsep}}
\renewcommand{\scriptsize}{\@setfontsize\scriptsize\@viiipt{9.5pt}}
\renewcommand{\tiny}{\@setfontsize\tiny\@vipt{7pt}}
\renewcommand{\large}{\@setfontsize\large\@xiipt{14pt}}
\renewcommand{\Large}{\@setfontsize\Large\@xivpt{18pt}}
\renewcommand{\LARGE}{\@setfontsize\LARGE\@xviipt{22pt}}
\renewcommand{\huge}{\@setfontsize\huge\@xxpt{25pt}}
\renewcommand{\Huge}{\@setfontsize\Huge\@xxvpt{30pt}}
\makeatother
\usepackage{epsfig}
\usepackage{times}
\usepackage{amsmath,amsfonts,amssymb,graphicx,graphics}
\usepackage{cite}
\usepackage[varg]{txfonts}
\usepackage{array}
\usepackage{multirow}
\usepackage{rotating}
\usepackage{setspace}
\usepackage{color}
 \usepackage{caption}
\usepackage{subcaption}
\usepackage{soul,color}
\usepackage{hhline}
\usepackage{footnote}
\usepackage[bottom]{footmisc}
\usepackage{setspace}
\usepackage{morefloats}
\usepackage{bibentry}
\usepackage{multicol}
\usepackage{fancyhdr}
\usepackage{notoccite}
\usepackage{listings}
\usepackage{algorithm, algpseudocode}
\usepackage{float}
\usepackage[hyphens]{url}
\usepackage{multirow, makecell}
\algrenewcommand\algorithmicindent{0.7em}%
\algblock[TryCatch]{Try}{EndTry}
\algrenewtext{Try}{\textbf{try}}
\algcblockdefx[TryCatch]{TryCatch}{Catch}{EndTry}
[1]{\textbf{catch} #1}
{\textbf{end try}}

\newcolumntype{C}[1]{>{\centering\let\newline\\\arraybackslash\hspace{0pt}}m{#1}}

\makeatother

\ssp

\begin{document}

% Declarations for Front Matter

\title{Novel Selectivity Estimation Strategy for Modern DBMS}
\author{Jun Hyung Shin}
\degreeyear{2018}
\degreesemester{Summer}
\degree{Master of Science}
% \chair{Professor Michael F. Modest}
% \othermembers{Professor Gerardo C. Diaz\\
% Professor Yanbao Ma}
% \numberofmembers{3}
% Committee members
\chair{Professor Florin Rusu}
\othermembers{Professor Stefano Carpin\\
Professor Mukesh Singhal}
\chairname{Professor Florin Rusu}
\othermembera{Professor Stefano Carpin}
\othermemberb{Professor Mukesh Singhal}
\numberofmembers{3}
\field{Electrical Engineering and Computer Science}
\campus{Merced}

\maketitle
\begin{frontmatter}
\copyrightpage
\approvalpage
\tableofcontents
\declarationpage
%\begin{dedication}  
%\null\vfil
%{\large
%\begin{center}
%To my parents
%\end{center}}
%\vfil\null
%\end{dedication}
\begin{acknowledgements}
 {\setstretch{1.0} I would like to thank Professor Florin Rusu for his guidance and encouragement that I learned so much about the database field from him. Without his effort, this work would not have been possible. I would also like to thank Professor Stefano Carpin and Mukesh Singhal for their guidance. Thanks to all the colleagues who I met here in UC Merced: Weijie Zhao, Yujing Ma, Ye Zhu, Tapish Rathore, Satvik Kul, Aditya Ranganath and many others. Thanks to my parents for their sincere support. Thanks to Jenny for her love and patience.}
\end{acknowledgements}
% \chapter*{Nomenclature and Abbreviations}
% {\setstretch{1.0} \input{nomenclature}}
\begin{abstract}   
\vspace{-15pt}
 \begin{center}
 \begin{minipage}{0.85\textwidth} 
\begin{center}
  \textbf{Novel Selectivity Estimation Strategy for Modern DBMS}\\~\\
 by\\~\\
 Jun Hyung Shin\\~\\
 Master of Science\\~\\
in\\~\\
Electrical Engineering and Computer Science\\~\\
University of California, Merced\\

 \end{center} 
 \end{minipage}
 \end{center}
\vspace{15pt}
\par
Selectivity estimation is important in query optimization, however accurate estimation is difficult when predicates are complex. Instead of existing database synopses and statistics not helpful for such cases, we introduce a new approach to compute the exact selectivity by running an aggregate query during the optimization phase. Exact selectivity can be achieved without significant overhead for in-memory and GPU-accelerated databases by adding extra query execution calls. We implement a selection push-down extension based on the novel selectivity estimation strategy in the MapD database system. Our approach records constant and less than 30 millisecond overheads in any circumstances while running on GPU. The novel strategy successfully generates better query execution plans which result in performance improvement up to 4.8$\times$ from TPC-H benchmark SF-50 queries and 7.3$\times$ from star schema benchmark SF-80 queries.
% \abstractsignature
\end{abstract}
\end{frontmatter}
\chapter{Introduction}\label{intro}
\section{Query Optimization}
Query optimization is an essential function in relational database systems to determine an optimal query execution plan among all possible plans that is estimated to be the cheapest so that a database system can use the least resources to run a query. To estimate the cost of each plan, a database system collects and maintains statistical information of stored data in system catalog to be used in optimization stage. The cost of executing each operation is then estimated based on the statistics and so as the cost of each plan. The plan with the minimal cost is determined to be an optimal query execution plan.

One of the major factors in the cost of query execution plans is the size of intermediate results since the resources used by a database system will increase when the system generates more intermediate outputs. Pushing operators, such as selections and projections, down below join operators and right above table scans is a simple but powerful tool to reduce the size of intermediate outputs generated by join operation. The cardinality of those pushed down operators should be estimated to examine the size of the intermediate output and the database statistics maintained by database systems can be used in this stage.

However, such cardinality estimation problem is still an open question that there are no solid solutions which can be applied to completely solve each and every case. A selection with complicate predicates and highly skewed or sparse data are the major factors of making this problem difficult. For instance, let us consider a query:
\begin{lstlisting}[caption={Selection with complicate predicates}, label=sql:complicate, basicstyle=\small\ttfamily]
SELECT R.C, R.D, S.E, T.F
FROM R, S, T
WHERE R.A = S.A AND S.B = T.B
AND R.A = x
AND R.B < y1 AND R.B > y2
AND (R.C = z1 OR R.C = z2);
\end{lstlisting}
The number of tuples in each relation $R$, $S$, and $T$ is 600 million, 5 million, and 1 million respectively, and primary keys in each relation, $S.A$ and $T.B$, are all distinct. Selectivity for a single attribute can easily be estimated if the data is uniformly distributed, and other statistical information, histograms for example, would provide more accurate estimation in case of skewed distribution of data. In this query, selectivity factor of each predicate $R.A = x$, $R.B < \textit{y1}$, $R.B > \textit{y2}$, and $(R.C = \textit{z1} \textit{ OR } R.C = \textit{u2})$ is estimated as 0.2, 0.167, 0.167, and 0.27 respectively. If we perform selection push-down over relation $R$ with those four predicates, the combined selectivity factor would be estimated as $0.2\times 0.167\times 0.167\times 0.27\approx 0.0015$ since they are all connected with \textit{AND} operator. In this case, the cardinality of $\sigma(R)$ is 0.9 million which now becomes the smallest among those relations. 

\begin{figure}[ht]
  \centering
  \begin{subfigure}{0.32\columnwidth}
    \includegraphics[width=\columnwidth]{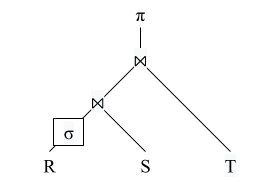}
    \caption{$(\sigma(R)\bowtie S)\bowtie T$}
  \end{subfigure}
  \begin{subfigure}{0.32\columnwidth}
    \includegraphics[width=\columnwidth]{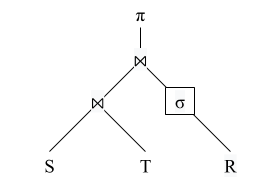}
    \caption{$(S\bowtie T)\bowtie \sigma(R)$}
  \end{subfigure}
  \begin{subfigure}{0.32\columnwidth}
    \includegraphics[width=\columnwidth]{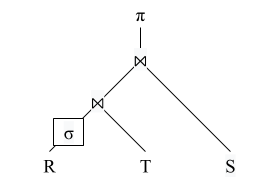}
    \caption{$(\sigma(R)\bowtie T)\bowtie S$}
  \end{subfigure}
  \caption{Possible query execution plans for query \ref{sql:complicate} with different join ordering}
  \label{fig:possible_plans}
\end{figure}

Determining the order of join operations is the next important step since the size of intermediate output highly depends on the join ordering. The size of intermediate output generated by a join operation can be estimated with this simplified formula:
\begin{displaymath}
|R\bowtie S|=\frac{|R||S|}{\max(\text{V}(R,R.A),\text{V}(S,S.A))},
\end{displaymath}
where $|R|$ is the number of tuples of relation $R$, $\text{V}(R,R.A)$ is the number of distinct values relation $R$ has in attribute $A$. Figure \ref{fig:possible_plans} shows three possible query execution plans in case of query \ref{sql:complicate} with different order of join operations. However, the third plan, $(\sigma(R)\bowtie T)\bowtie S$, would be very expensive to execute since it has to perform Cartesian product between $\sigma(R)$ and $T$ due to the fact that there is no join condition between those two relations. Hence, let us compare the size of intermediate output from the first two plans. The size of intermediate output generated by the first plan, $(\sigma(R)\bowtie S)\bowtie T$, is 0.9 million, while the one generated by the second plan, $(T\bowtie S)\bowtie \sigma(R)$, is 5 million. Hence, the system would choose the first plan $(\sigma(R)\bowtie S)\bowtie T$ because the size of intermediate output generated by this query execution plan is less than the second plan generates. In fact, however, the actual selectivity factor of $\sigma(R)$ stays at 0.167 due to huge overlaps between the four predicates, which means the estimated selectivity factor is erroneous; this is where selectivity estimation problem begins. With the actual selectivity, the size of intermediate output generated by the first plan becomes 100.2 million, while the second plan still generates 5 million. Therefore, the actual optimal order of join would be $(T\bowtie S)\bowtie \sigma(R)$ that can really minimize the size of intermediate outputs, instead of the chosen $(\sigma(R)\bowtie S)\bowtie T$. In this specific case, the size of intermediate output generated by the chosen plan is 20 times larger than the one from the optimal plan. The gap between those two plans could be much larger once the size of relation increases, which results in performance degradation by choosing a sub-optimal plan due to the wrong selectivity estimation. Estimation would be more erroneous when the selection contains more complicate conditions such as \textit{OR} and \textit{LIKE}. We therefore need a better selectivity estimation strategy which can cover such complicate predicates to avoid choosing a bad query execution plan.

\section{Modern Database Systems}
While this problem is widely studied for decades, in the meantime, database systems have been evolved along with the evolution of hardwares. Over the past decade, GPU (Graphics Processing Unit) has been utilized to accelerate database systems. Although GPU is used primarily for 3D image processing, it also has been widely used for accelerating systems recently due its massively parallel architecture; unlike CPU (Central Processing Unit), which has just small number of cores with large cache memory to deal with a few threads at the same time, a GPU contains thousands of cores with smaller cache memory, so that it is more suitable for executing huge but simple batches of data since it can handle thousands of threads simultaneously \cite{NVIDIA:2009}. As the volume of data has become drastically larger, GPU can help database systems to process huge amount data in parallel to achieve better performance.

GPU-accelerated relational database systems use main memories as their primary data storage instead of disk-based second storage devices, HDD (Hard Disk Drive) for example. GPU cannot be fully utilized under disk-based storage systems because fetching data from this kind of storages to GPU takes too much time due to the I/O bottleneck \cite{He:2009:RQC:1620585.1620588}; the bandwidth of a typical HDD is only a few hundreds of MB/s while a DRAM can process up to tens of GB/s. 

\begin{table}[ht]
  \centering
  \caption{Column vs. row stores}
  \label{table:col_vs_row}
  \begin{tabular}{|r|c|l|l|r|} \hline
    rowid & empid & name\_last & name\_first & salary \\ \hline
    1 & 111 & Rogers & Steve & 8500\\ \hline
    2 & 112 & Romanoff & Natasha & 10700 \\ \hline
    3 & 113 & Banner & Bruce & 2900\\ \hline
    4 & 114 & Strange & Stephen & 5300\\ \hline
  \end{tabular}
  \par
  ~\newline
  \par
  \begin{tabular}{l|l}
    \multirowcell{4}{Column\\store} & 111:1,112:2,113:3,114:4;\\
    & Rogers:1,Romanoff:2,Banner:3,Strange:4;\\
    & Steve:1,Natasha:2,Bruce:3,Stephen:4;\\
    & \hl{8500:1,10700:2,2900:3,5300:4;}\\ \hline
    \multirowcell{4}{Row\\store} & 1:111,Rogers,Steve,\hl{8500};\\
    & 2:112,Romanoff,Natasha,\hl{10700};\\
    & 3:113,Banner,Bruce,\hl{2900};\\
    & 4:114,Strange,Stephen,\hl{5300};\\
  \end{tabular}
\end{table}

The storage layout is column-oriented that stores data by column rather than by row. Table \ref{table:col_vs_row} shows an example of how data is stored in and read from column and row-oriented systems; when a user tries to read salary, column-oriented system can read only the necessary columns, the last column which is consist of salaries highlighted in the figure, while row-oriented system still need to read all the rows first and then extract the salary from each row to perform the same task. Therefore, column-oriented systems use less I/O and transfer less data between CPU and GPU. Moreover, by storing data in columnar layout, it can have uniform data type for each column that can achieve higher compression rate that is not available in row-oriented system. Data transfer between host and GPU memory is critical and column-oriented system can help reducing the amount of data transferred, therefore it indeed results in faster execution time and less memory requirement compare to row-oriented system, especially when the system has huge datasets \cite{Bress:2014}.

The processing model of GPU-accelerated database systems is operator-at-a-time rather than tuple-at-a-time. Operator-at-a-time approach is more cache friendly due to its bulk processing characteristic, but requires more memory to store the intermediate results, while tuple-at-a-time model only generates tiny intermediate data. However, tuple-at-a-time approach needs virtual function calls to process each tuple for required operators and GPU is not capable of running such complex function calls \cite{Graefe:1990:EPV:93597.98720}.

\section{Novel Approach and Contributions}
Back to the selectivity estimation problem, we propose a novel approach which can guarantee to have the exact cardinality of a selection to prevent such a worst case scenario. The key idea of this approach is running an extra aggregate query to compute the selectivity of a selection during the query optimization phase that can be used further to estimate the cost of query execution plans.
\begin{figure}[ht]
  \centering
  \begin{subfigure}{0.67\columnwidth}
    \includegraphics[width=\columnwidth]{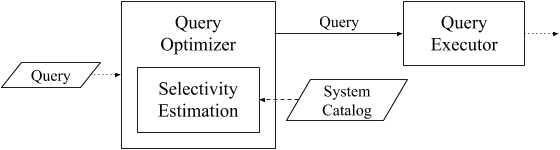}
    \caption{Traditional approach}
  \end{subfigure}
  \par\medskip
  \begin{subfigure}{0.67\columnwidth}
    \includegraphics[width=\columnwidth]{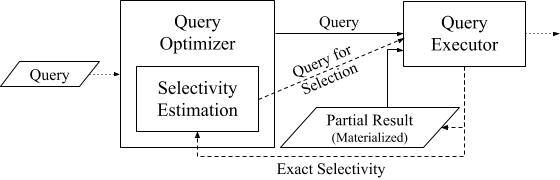}
    \caption{Novel approach}
  \end{subfigure}
  \caption{Traditional vs. novel selectivity estimation}
  \label{fig:old_vs_new}
\end{figure}
As illustrated in figure \ref{fig:old_vs_new}, our approach interacts with query executor during the query optimization phase for selectivity estimation process, while traditional approach only accesses system catalog to utilize any synopses or statistics which does not fit well into the selection with complex predicates. We pass a partial query which corresponds to the selection we want to evaluate to the query executor, and manipulate the partial query a bit to compute the exact selectivity that will be returned to the query optimizer. Plus, we materialize the partial result from the query executor so that we can simply reuse the intermediate result set instead of reprocessing the same selection during the full query execution.

While this strategy can compute and return the exact selectivity, it might cause serious overheads due to I/O access time; for each time we pass the partial query to the executor, the system will read data storage and accessing disk-based second storage devices will become the bottleneck. However, the evolution of database systems, primary data storage is replaced by main memory and accelerated by GPU, has significantly decreased the response time, and this is the reason why we can propose the novel approach at this moment that now becomes a feasible solution.

\begin{itemize}
  \item We show a way to successfully estimate selectivity regardless of complexity in predicates and data distribution that is not always possible with the existing synopses and statistics.
  \item Our approach does not rely on any database synopses and statistics and rather simple to apply for any modern database systems.
  \item We implement selection push-down extension for a GPU-accelerated relational database system, MapD, that pushes only effective selections down based on the estimated selectivity from our strategy.
  \item We run experiments over synthetic dataset to show that our approach does not involve serious overhead, less than 30 ms, in any circumstance and successfully boosts the performance up to 4.8$\times$ from TPC-H benchmark queries in scale factor of 50 and 7.3$\times$ from SSB queries in scale factor of 80.
\end{itemize}

\pagebreak
\chapter{Selectivity Estimation}\label{preliminary}
Selectivity estimation problem has been studied for decades, hence there are numerous approaches for this single step to determine an optimal query execution plan. Beginning from simple heuristics, for example, the cardinality of a selection which has a single condition that an attribute is equal to a constant can be estimated by using this simplified formula:
\begin{displaymath}
|\sigma_{A=x}(R)|=\frac{|R|}{\text{V}(R,A)}.
\end{displaymath}
In case of an inequality predicate, this formula can be used:
\begin{displaymath}
|\sigma_{A<x}(R)|=\frac{|R|}{3},
\end{displaymath}
where the number of tuples divided by three is just a heuristic function that such an inequality predicate tends to retrieve a small fraction of the possible tuples \cite{Garcia-Molina:2008:DSC:1450931}.

In the real world, however, multiple predicates are quite common as shown in an example query \ref{sql:complicate}. This is a query with multiple predicates for relation $R$, while other selection predicates are abbreviated. If we perform selection push-down for relation $R$, we need to consider three columns $A$, $B$, and $C$ simultaneously. $1/\text{V}(R,A)$ and $1/3$ are selectivity factors for equality and inequality predicates respectively and can be multiplied to $T(R)$ in case of multiple predicates connected by \textit{AND}. In case of predicates connected by \textit{OR}, we can simply assume that they are independent and subtract joint probability from the sum of all selectivity factors for predicates. Therefore, the cardinality of this selection can be estimated as:
\begin{align*}
|\sigma&_{R.A = x\textit{ AND }R.B < y1\textit{ AND }R.B > y2\textit{ AND }(R.C = \textit{z1}\textit{ OR }R.C = \textit{z2})}(R)| = \\
&|R|\times \frac{1}{\text{V}(R,A)}\times \frac{1}{3}\times  \frac{1}{3}\times\left(\frac{1}{\text{V}(R,C)}+\frac{1}{\text{V}(R,C)}-\left(\frac{1}{\text{V}(R,C)}\right)^{2}\right)
\end{align*}
This approach has assumptions that the data distribution is uniform and the output from each predicate is independent. These assumptions are unfortunately not realistic in real world datasets that even skewed and sparse distribution of data is very popular. In such cases, the estimation can be wrong by orders of magnitude and may result in a bad query execution plan which makes the system use unnecessary resources.

\begin{figure}[ht]
  \centering
  \begin{subfigure}{0.49\columnwidth}
    \includegraphics[width=\linewidth]{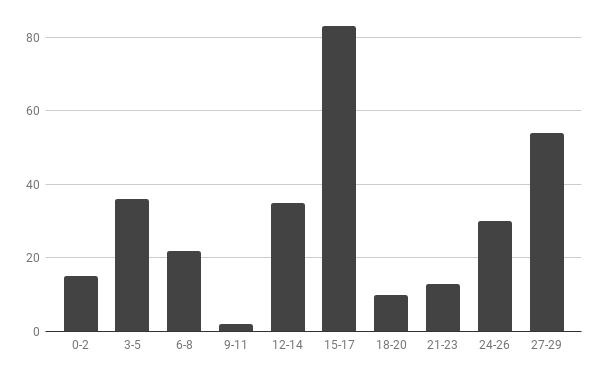}
    \caption{Equi-width histogram}
  \end{subfigure}
  \begin{subfigure}{0.49\columnwidth}
    \includegraphics[width=\linewidth]{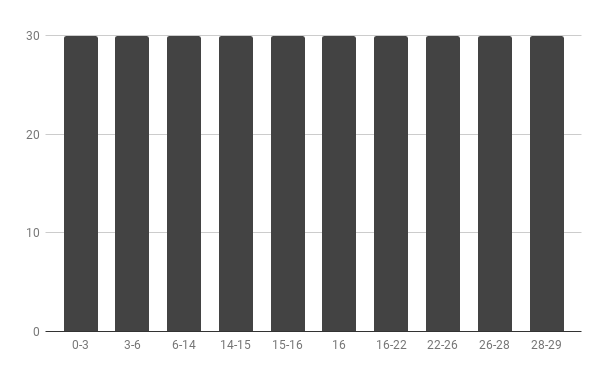}
    \caption{Equi-depth histogram}
  \end{subfigure}
  \caption{Examples of equi-width and equi-depth histograms}
  \label{fig:hist}
\end{figure}f
To provide more accurate cardinality estimation, numerous relational database systems use histograms to store the column statistics that values in a column are distributed into a certain amount of buckets in a histogram. Since Histograms are relatively easy to build and interpret but capable of represent data distribution accurately, they are broadly studied and applied to plenty of relational database systems. The most basic histograms are \textit{equi-width} and \textit{equi-depth}, where the latter were proposed later and to better adapt to skewed distribution of data., as shown in figure \ref{fig:hist}, in case of equi-depth histograms, for example, every bucket in a histogram contains the same number of data, i.e. $D=T(R)/B$ where $D$ is the depth and $B$ is the number of buckets. The selectivity for an equality predicate can be estimated by this simplified formula:
\begin{displaymath}
|\sigma_{A=x}(R)|=\frac{D}{\text{V}(b_{x})},
\end{displaymath}
where $\text{V}(b_{x})$ is the number of distinct values in the bucket that contains x. For example, in figure \ref{fig:hist}, if we try to estimate the selectivity where the attribute involved equals to 16, $30/2 + 30/1 + 30/7 = 49.3$ would be the returned. This approach however requires an assumption that the data distribution within a bucket is uniform, therefore many alternatives such as \textit{max-diff}, \textit{compressed}, and \textit{v-optimal} histograms have been introduced to store the column statistics into the limited number of buckets for more accurate cardinality estimation. Meanwhile, these types of histograms have limitation on queries with multiple predicates since each histogram provides statistics of only a single column and there are no correlations between attributes. Multi-dimensional histograms can be alternatives but also involve technical challenges because the correlations become exponentially more complex in case of those queries, which are quite common nowadays, thus system resource usage significantly increases \cite{Ioannidis:2003:HH:1315451.1315455}.

\begin{figure}[h]
  \centering
  \includegraphics[width=\linewidth]{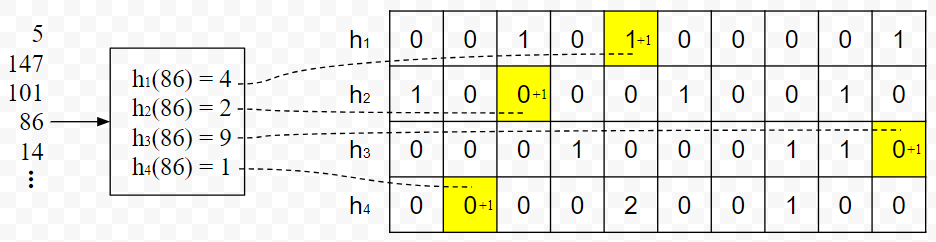}
  \caption{Examples of count-min Sketch}
  \label{fig:cm_sketch}
\end{figure}
\pagebreak
Sketches are another options that are a type of data streaming summary. This approach basically maintains a summary of updates so far, and this summary is modified by each update in a stream that is irrespective of previous updates so that the summary can be used for estimation instead of full dataset at any time. More specifically, a sketch involves a linear transform where a fixed-size sketch matrix is multiplied by data as a column vector to create a sketch in a form of vector and such a transform is based on simple and fast hash functions which are suited for processing data streams. A sketch is designed for answering a particular set of queries by performing a specific procedure over the sketch. Frequency based sketches such as \textit{count-min}, \textit{count}, and \textit{AMS} sketches can be used for such estimation problems. For example, figure \ref{fig:cm_sketch} illustrates how a sketch vector is updated from an input stream. In this case, it has four hash functions which corresponds to each row of the vector. The output from each hash function corresponds to the column for each row of the sketch vector and adds 1 to the counter. When we estimate selectivity of 86, we simply get the minimum counter value corresponding to the output of $h_i(86)$ for each row. However, sketches also have difficulties on high-dimensional queries which exponentially increases the size complexity \cite{Cormode:2012:SMD:2344400.2344401}.

Sampling can be used for cardinality estimation that randomly draws data from a dataset by a certain size and then scales up the selection from the sample by a factor of the sample size. The selectivity for any predicate can be estimated by this simplified formula:
\begin{displaymath}
|\sigma_{A<x}(R)|=|\sigma_{A<x}(R')|\times \frac{|R|}{|R'|},
\end{displaymath}
where $R'$ is a sample randomly drawn from $R$. It is also supported by numerous database systems and a variety of techniques have been studied like histograms, but has several advantages over histograms. First of all, samples are not affected by the number of attributes in selection, i.e. dimension, that no matter how many attributes are involved in a selection, all we do is drawing some random data and scaling up afterward. Therefore, this approach is much more applicable to high-dimensional query which is a major problem of histograms. Plus, it generally does not depend on pre-built statistics model that database systems do not have to collect and maintain column statistics, but just runs on the go. On the other hand, sampling is more suitable for detecting the broad patterns but less suitable for queries which select only a single or a few tuples due to its random characteristics that such rare cases are unlikely drawn in a sample and results in a wrong estimation. In the same context, it is generally sensitive to skewed and sparse data that sampling is not a better approach for selectivity estimation since such cases are so popular in real world.

LEO \cite{Markl:2003:LAQ:1014767.1014781} is another type of query optimizer that is an alternative to overcome the selectivity estimation problems with complex predicates. It basically monitors queries during query execution to retrieve the actual cardinality of each operation in the query execution plan. Then, it compares the actual cardinalities with previous estimates to check whether the estimations were correct. If the estimates were erroneous, the autonomic optimizer computes adjustments to the estimates so that it can be used for the similar queries in the future. Moreover, it can perform query re-optimization \cite{Kabra:1998:EMR:276304.276315} in case of notable error in the estimates to fix the plan derived from the estimates. Nevertheless, such mid-query re-optimization involves overhead to build new plan which discards pre-constructed structures such as hash tables for the previous plan. Also, it would not be useful if the plan generated by the estimates is already optimal since it only involves overhead to collect the statistics for operators. Plus, as long as the query optimizer depends on the database statistics and synopses, there is no improvement for the queries which were not anticipated.
\pagebreak
\chapter{A Novel Query Optimization Paradigm}\label{new_approach}
\begin{figure}[ht]
  \centering
  \includegraphics[width=0.5\columnwidth]{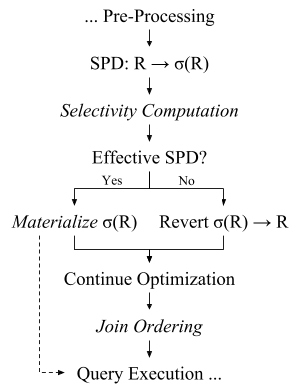}
  \caption{Selection push-down decision process}
  \label{fig:spd_decision}
\end{figure}
In this section, we present the novel selectivity estimation and further optimization strategy. Figure \ref{fig:spd_decision} illustrates how this strategy basically works on a relation \textit{R}. The query optimizer first performs selection push-down for relation \textit{R}, and \textit{computes} the exact selectivity from $\sigma(R)$. The selection is now evaluated based on the computed selectivity to determine whether it is effective selection push-down or not; its impact on \textit{join ordering} which can reduce the size of intermediate output from join operations, and \textit{materialization} of $\sigma(R)$ which involves extra I/O that should be minimized are considered. Once the pushed down selection is turned out to be effective, the system materializes the result set of $\sigma(R)$ so that it can use it during the full query execution. Otherwise, the push-down will be reverted. After processing all the other relations, the system performs join ordering based on the computed selectivities so that it can execute the given query with an optimal query execution plan in the end.

\section{Selectivity Computation}\label{selectivity_computation}
\begin{figure}[h]
  \centering
  \includegraphics[width=0.67\columnwidth]{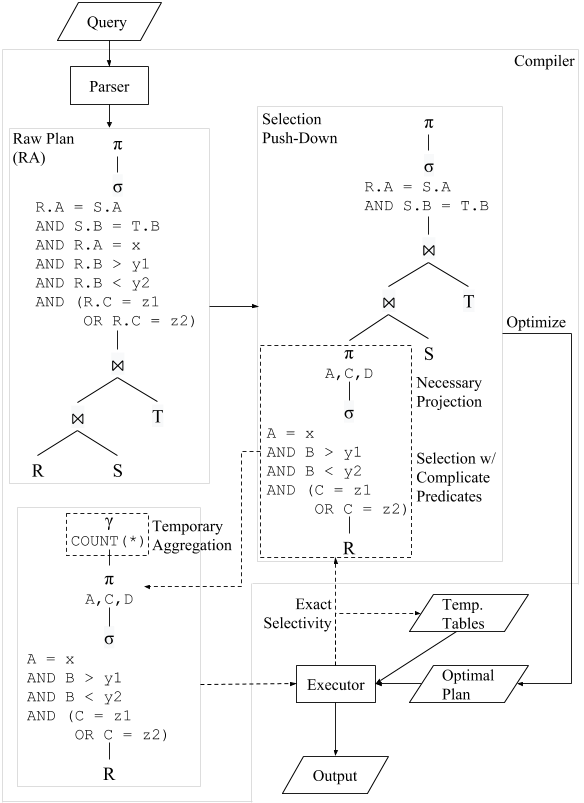}
  \caption{Flowchart of selectivity estimation for relation R by running an extra aggregate query inside query compiler}
  \label{fig:flowchart}
\end{figure}
Figure \ref{fig:flowchart} shows a brief overview of how selectivity estimation is performed from the query \ref{sql:complicate} with our approach and how query is executed. Once the query is passed to the query compiler, it is parsed into a raw plan where it is a simple relational algebra tree that compiler can understand. In the query optimization stage, the system searches for selection predicates and determines whether they can be pushed down in the relational algebra tree as a preliminary step. This is a typical selection push-down, a simple but most effective way to reduce the size of intermediate output generated by join operators since the size of intermediate output is basically proportional to the product of the size of two input relations. Therefore, the size of this intermediate output can be reduced by pushing corresponding selection, $R.A = x\textit{ AND }R.B < \textit{y1}\textit{ AND }R.B > \textit{y2}\textit{ AND }(R.C = \textit{z1}\textit{ OR }R.C = \textit{z2})$, down for relation \textit{R} based on the given query. 

At this point, the cardinality of relation \textit{R} including pushed down selection should be estimated so that the system can determine the optimal way to join relations involved in the given query. The selectivity estimation is however problematic in case of complicate predicates like the given query as previously shown, and may lead to a bad query execution plan if it is erroneous. Then what method can we use to avoid such problematic situations? One idea is that, if we simply focus on the accuracy of the selectivity, we can run a query which has a simple aggregation, \textit{COUNT}, \textit{FROM} the relevant table, and corresponding selection predicates in \textit{WHERE} clause. From the given query \ref{sql:complicate}, we can run a query to estimate selectivity of selection for relation \textit{R}:
\begin{lstlisting}[caption={Selectivity computation}, label=sql:selectivity, basicstyle=\small\ttfamily]
SELECT COUNT(*) 
FROM R
WHERE A = x
AND B < y1 AND B > y2
AND (C = z1 OR C = z2);
\end{lstlisting}
The output of query \ref{sql:selectivity} will be the exact cardinality of the given selection, and this is where our strategy begins; we execute a simple aggregate query of \textit{COUNT} for the selection with such complex predicates over relation \textit{R} to compute the selectivity.

There are a few steps to generate the query above from the full relational algebra tree. The system first pushes not only the selection but also a projection down. Although this is not necessary for the selectivity computation itself, it can help reducing I/O access because typical GPU-accelerated relational database systems store the data in columnar layout. If no column is specified with a projection, the query executor will read the whole column and this will involve unnecessary overhead. Hence, we need to look up all the predicates to find the columns which are necessary for the full query processing and push all of them down. In this case, the projection contains column \textit{A}, required for join operation, \textit{C} and \textit{D}, projected at the top the relational algebra tree. We will discuss more about this in Section \ref{materialization}. 

After pushing operators down, the sub-tree we have now is just a simple \textit{SELECT-FROM-WHERE} query. To compute the cardinality of this, the system adds a temporary aggregate operator which has \textit{COUNT} function at the top of the sub-tree, as shown in bottom-left of figure \ref{fig:flowchart} to generate a relational algebra tree which now fully represents the query \ref{sql:selectivity}. At this point, the system now can pass the sub-tree with the temporary aggregation to the query executor, instead of referencing any synopses or statistics. Once the query executor runs the sub-tree, the exact selectivity will be returned that is the key of join ordering process.

\section{Join Ordering}\label{join_ordering}
An optimal query execution plan is the plan with the least cost to process the query. The cost is highly affected by the size of intermediate outputs mostly generated by join operations. Therefore, join ordering is the most significant step to find an optimal plan. Since the selectivities computed in the previous step are accurate, we can estimate the size of each possible intermediate output more precisely and find an optimal order of join operations for sure. We provide some examples to discuss how selection push-down affects the cost of join operations and determine whether it is effective or not.

\begin{figure}[ht]
  \centering
  \begin{subfigure}{0.33\columnwidth}
    \includegraphics[width=\linewidth]{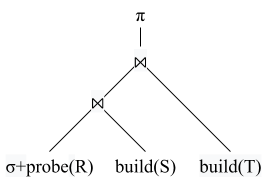}
    \caption{Naive plan}
    \label{fig:plan_probe_naive}
  \end{subfigure}
  \par
  \begin{subfigure}{0.33\columnwidth}
    \includegraphics[width=\linewidth]{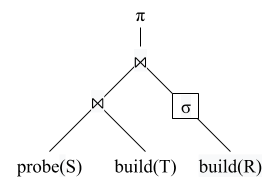}
    \caption{Order of joins changed}
    \label{fig:plan_probe_fpd_true}    
  \end{subfigure}
  \begin{subfigure}{0.33\columnwidth}
    \includegraphics[width=\linewidth]{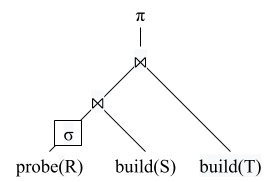}
    \caption{Order of joins not changed}
    \label{fig:plan_probe_fpd_false}    
  \end{subfigure}
  \caption{Selection push-down on the relation to be probed}
  \label{fig:plan_probe}
\end{figure}

First of all, let us take a look at the naive plan which does not have any pushed down selections, shown in figure \ref{fig:plan_probe_naive} which is derived from the query \ref{sql:complicate} to get the basic idea of how the system choose an optimal order of join operations; assume that $|R|>|S|>|T|$ and each tuple in the join column is distinct. The query execution plan has a typical left-deep join tree that probes over the largest relation \textit{R} and builds hash tables for other relations, \textit{S} and \textit{T}, to perform a hash join. The reason why we probe on the largest relation is because the memory capacity is quite limited, specifically in GPU which only has a few GB for a single device, and even if we have enough memory space, building a hash table for the larger relation would be more expensive. The order of joining other relations are based on two major factors. Cardinality of each relation is important since joining smaller relation first would reduce much more tuples ahead of time. To prevent a Cartesian product instead of any conditional join, the presence of join predicates with the probed relations is also significant; this is why the naive plan joins relation \textit{R} and \textit{S} first even if the cardinality of \textit{T} is the smallest. All selection predicates will be evaluated while probing since no selection is pushed down. Before the physical query execution, code generation techniques will be applied to combine all those join operations to perform all of them in a single pipeline so that it does not materialize any intermediate output during execution. 

A selection push-down indeed reduces the cardinality of corresponding relation, but it also involves intermediate output which should be materialized in memory that extra times to write and read materialized result set are required to begin the pipeline processing. Therefore, we need to examine whether such selection push-down is actually helpful or not against the materialization. If the materialization is determined as expensive, we should discard the pushed down selection to avoid unnecessary overhead.

We can first think of some cases when the selection push-down is on the relation to be \textit{probed}, \textit{R} in this case. As a starter, let us consider a case when the cardinality of relation \textit{R} becomes smaller than the second largest relation \textit{S} by selection push-down, $|S|>|\sigma(R)|>|T|$. This actually changes the order of join operations as illustrated in figure \ref{fig:plan_probe_fpd_true}. To compare the cost of join operations from each plan, we use the equation in Chapter \ref{intro} to estimate the size of intermediate output. For example, the size of intermediate output in the figure \ref{fig:plan_probe_fpd_true} that is from $S\bowtie T$ can be estimated as:
\begin{align*}
&\frac{|S||T|}{\max(\text{V}(S,B),\text{V}(T,B))}=|S| &(\because \text{V}(S,B)=\text{V}(T,B)=|T|)
\end{align*}
In the naive plan, the intermediate output is generated by $R\bowtie S$. However, we do not necessarily join the whole tuples in relation \textit{R} because we can join only tuples which satisfy the conditions by evaluating selection predicates before we actually probe each tuple. For this reason, the size of intermediate output in the naive plan would be estimated as $\sigma(R)$ hence the size of intermediate output actually increases by pushing selection down. The selection push-down also involves extra times for writing and reading materialized $\sigma(R)$ which could be expensive if the selectivity of $\sigma(R)$ is high.

If $|\sigma(R)|>|S|>|T|$, then the order of join operations does not change as shown in figure \ref{fig:plan_probe_fpd_false}, which actually becomes worse. The size of intermediate output would be equal, including the fact that we build hash tables for the same relations, but this also involves extra materialization. For these reasons, selection push-down for the relation to be probed, typically the largest among them, cannot be considered as effective approach, rather bring extra overhead due to materialization, that we should avoid.

\begin{lstlisting}[caption={Selection on a build input}, label=sql:build, basicstyle=\small\ttfamily]
SELECT R.C, R.D, S.E, T.F
FROM R, S, T
WHERE R.A = S.A AND R.B = T.B
AND S.A = x
AND S.B < y1 AND S.B > y2
AND (S.C = z1 OR S.C = z2);
\end{lstlisting}

\begin{figure}[ht]
  \centering
  \begin{subfigure}{0.33\columnwidth}
    \includegraphics[width=\linewidth]{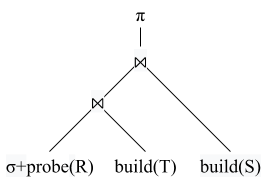}
    \caption{Naive plan}
    \label{fig:plan_build_naive}
  \end{subfigure}
  \par
  \begin{subfigure}{0.33\columnwidth}
    \includegraphics[width=\linewidth]{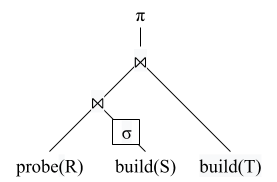}
    \caption{Join ordering changed}
    \label{fig:plan_build_fpd_true}    
  \end{subfigure}
  \begin{subfigure}{0.33\columnwidth}
    \includegraphics[width=\linewidth]{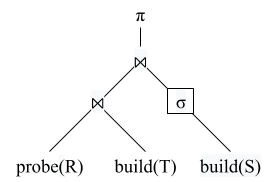}
    \caption{Join ordering not changed}
    \label{fig:plan_build_fpd_false}    
  \end{subfigure}
  \caption{Selection push-down on the relation as a build input}
  \label{fig:plan_build}
\end{figure}

Let us consider other cases based on the query \ref{sql:build} and assume all assumptions still hold. In this case, selection is on relation \textit{S} which is now a \textit{build} input. Since a join condition is changed, the naive plan is also affected as shown in figure \ref{fig:plan_build_naive} which joins relation \textit{R} and \textit{T} first to minimize the intermediate output. Selection predicates are again evaluated while probing since selection is not pushed down, which means the hash table for relation \textit{S} contains the whole tuples with all necessary columns and this gives us more room to improve. 

In case of the cardinality of relation \textit{S} becomes the smallest after selection push-down, $|R|>|T|>|\sigma(S)|$, which changes the order of join operations as illustrated in figure \ref{fig:plan_build_fpd_true}. The size of intermediate output would be reduced from $|R|$ to $|\sigma(S)|$ and we can also reduce the cost of join operation by the factor of $|S|-|\sigma(S)|$ by building a smaller hash table based on filtered \textit{S}. This is because we first join every tuple in \textit{S} to evaluate the selection predicates in figure \ref{fig:plan_build_naive}, while we join only the tuples in \textit{S} which satisfy the condition in figure \ref{fig:plan_build_fpd_true}. The expected overheads are again from writing and reading the materialized result set generated by $\sigma(S)$ that we always need to consider.

On the other hand, if $|R|>|\sigma(S)|>|T|$ that the selectivity of $\sigma(S)$ is very high, the order of join operations would not be changed as shown in figure \ref{fig:plan_build_fpd_false}. In this case, the size of hash table for relation \textit{S} is reduced by $|S|-|\sigma(S)|$ that can directly affect the cost of join operation. If the cost we can save is greater than the cost of processing materialized result set, the pushed down selection can be considered as an effective operation. Therefore, the effectiveness of the selection push-down is very sensitive to the selectivity of $\sigma(S)$; if the selectivity is too high, the overheads from writing and reading materialized result set would overwhelm the profit of reduced cost of join operations.

This approach shares the idea of modifying the query execution plan based on the estimated selectivity with query re-optimization. Query re-optimization performs join ordering with estimated selectivities first, and  runs the plan to collect statistics of selections with extra operators to get more accurate selectivity. If the estimated selectivity is turned out to be erroneous, the system performs join ordering again based on the new selectivities to find more optimal plan. In our approach, the system can compute the exact selectivities without introducing extra operators in the query execution plan and decide whether a selection is worth to push down or not by foreseeing the impact on join ordering based on the selectivities. An optimal join ordering can be determined consequently before starting the full query execution.

\section{Materialization}\label{materialization}
As we discussed, processing materialized data causes extra times to write and read. The overall performance would be degraded when the size of materialized data becomes huge even if the cost of join operations is reduced by selection push-down. Therefore, we need to minimize the size of materialization as much as we can to reduce the unnecessary overhead.

Let us first think of how the size of a relation affects the effectiveness of selection push-down. Suppose we have a selection which has 0.1 selectivity for a relation \textit{R}. If the size of \textit{R} is one thousand for example, 900 tuples will be filtered out. This number, $|R|-|\sigma(R)|$, is the actual advantage that we can take from the selection push-down since we only need to materialize 100 tuples instead of the whole tuples. If the size of the relation is one million, 900 thousand tuples will be filtered out in this case. The latter reduces relatively much more tuples than the former does hence we can say that the effectiveness of selection push-down increases as the size of the corresponding relation increases. Therefore, if the size of a relation is too small, the advantage we can take from selection push-down also becomes tiny that such selection push-down can be worthless. The small advantage of selection push-down on small relation can be offset by extra costs for processing not only the materialized data but also selection push-down and selectivity computation even though they are small. If the size of relation is large enough, we need to look up the largest relation so that we can avoid unnecessary materialization because selection push-down on the relation to be probed is turned out to be inefficient.

To do this, we let the system collect meta data, specifically the size, for each relation involved in the given query and construct a priority queue of relations sorted by size in descending order at the beginning of the query optimization stage. While inserting each relation into the queue, we introduce a threshold to discard relations smaller than the threshold so that we do not spend extra times to evaluate selections on those small relations which are unlikely the effective operation. Once the queue is filled only with large enough relations, we begin to evaluate the selection of each relation in the priority queue by computing its selectivity from the second relation to skip the largest relation to be probed. Profit and loss for each selection push-down should be compared based on the cost model or heuristics to avoid huge materialization and figure out whether each selection can bring performance improvement.

\begin{figure}[h]
  \centering
  \includegraphics[width=\columnwidth]{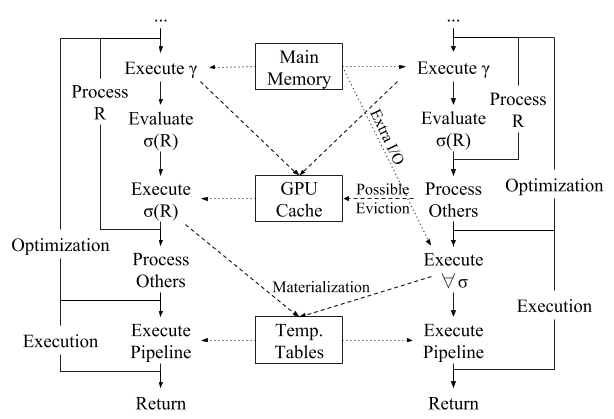}
  \caption{Materialize $\sigma(R)$ during optimization phase (left) vs. during execution phase (right)}
  \label{fig:when_to_materialize}
\end{figure}

On the other hand, we can think of when to materialize the intermediate output. When a selection is pushed down, it eventually generates an intermediate output which should be materialized at some point. In our approach, as shown in the left side of the figure \ref{fig:when_to_materialize}, once the system determines $\sigma(R)$ is worth to push down based on its selectivity and the cost model, it executes the sub-tree by removing the temporary aggregate which used for selectivity computation to generate the result set for $\sigma(R)$. By materializing the selection during the query optimization phase, we can utilize the scan of relation \textit{R} which is already cached into GPU during selectivity computation so that we do not have scan again from the storage. The result set will be materialized as a temporary table and used for the full query execution. Back in selectivity computation phase, we mentioned that a projection is also pushed down along with the selection. This Projection push-down is necessary for this step in order to read only the necessary columns so that we can minimize the size of materialized result set. If we do not materialize right after the selectivity computation, the intermediate output would be materialized right before processing the pipeline during the query execution phase as shown in the right side of the figure \ref{fig:when_to_materialize}. In that case, we cannot guarantee that the scan of relation \textit{R} is still cached in GPU since it could be evicted while processing other relations during query optimization phase. If the scan of relation \textit{R} is evicted, the query executor again has to read the storage to retrieve the data from \textit{R} that involves extra I/O access. Therefore, we materialize each selection right after the selectivity computation for each to fully utilize the cached data.

\section{Advantages and Concerns}
The novel selectivity estimation strategy has several strengths over the existing statistical approaches:
\begin{itemize}
\item \textit{Accurate}: Needless to say, the core of this approach comes from the exact selectivity by running an extra aggregate query on demand. It is actually no longer an estimation based on any kind of statistical models, but a computation of what the database system currently stores. It is obvious that the accuracy of selectivity is no longer affected by the number of attributes involved which is one of the major factors of wrong estimation in histograms and sketches. Moreover, the selectivity returned is not only accurate but also consistent no matter how frequent the database is updated, while the other existing statistic models require extra maintenance to fully reflect the updates.
\item \textit{Comprehensible}: Variety of statistical methods for more accurate and efficient selectivity estimation have been studied for decades. However, further understanding about the statistical backgrounds is necessary to apply and extend those ideas for database systems. On the other hand, our strategy is rather simple that it does not require any complicate synopses or statistical comprehension for the estimation process. All we need is simply building an aggregate query and its execution to retrieve the exact selectivity.
\item \textit{Easy to apply}: In perspective of the actual implementation, only a few steps are required to compute the selectivity. We first need to modify the query execution plan a little bit by adding a simple and consistent, always a \textit{COUNT}, aggregate operator and connect it to the pushed down selection as an extra query generation. We can simply reuse the same executor function calls for the query execution part that can be done by simply passing the newly generated plan where the root node is the aggregate operator. While there is nothing new or complex processes to be written in this strategy, the existing statistical approaches require their own new functions or interfaces for the estimation.
\end{itemize}

Nevertheless, the cost of running a query to retrieve the exact selectivity would be expensive than the existing strategies based on statistics and synopses. Space cost would not be much because \textit{COUNT} is the simplest one-pass aggregation function which only requires one small block to accumulate the total count so far, but the time cost would be relatively much more. Running aggregate queries will eventually cause some overheads and if the sum of overheads is greater than the improvement by choosing a better query execution plan, this approach would not make any sense. Since the overheads are mostly due to I/O access, they are actually highly dependent on the architectures of a database system; our approach would not work well on any disk-based database systems for example. This is why our strategy becomes feasible at this time since the rise of in-memory and GPU-accelerated database systems. We will discuss further detail in chapter \ref{experiments} based on our experiments.
\pagebreak
\chapter{Implementation}\label{implementation}
Based on the new selectivity estimation strategy covered in the previous chapter, we implement a selection push-down extension for \textit{MapD}, a GPU-accelerated database system. We will briefly introduce about MapD and more detail of our implementation for the optimization next.

\section{GPU-Accelerated DBMS}\label{mapd_intro}
MapD is a platform for real-time visual analytics on massive datasets based on its own core database engine, MapD Core. The core is an open source relational database system designed for running on GPU. MapD Core uses main memory as its primary storage, stores data in columnar layout, and adopts operator-at-a-time as its processing model to achieve the best performance.

\begin{figure}[h]
  \centering
  \includegraphics[width=\columnwidth]{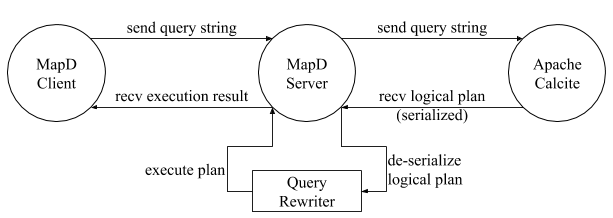}
  \caption{Query Flow in MapD Core}
  \label{fig:mapd_flow}
\end{figure}

Figure \ref{fig:mapd_flow} illustrates how a query is processed in MapD Core, specifically in the sense of query compilation. Once a query string in SQL is passed from MapD client to server, it directly passes it to Apache Calcite, a data management framework that provides standard functionalities to process SQL, that parses the raw query string and generates raw logical query execution plan in relational algebra in serialized JSON. The raw plan is passed back to MapD server and it deserializes the raw plan in the format that it can understand. After the deserialization, the logical plan advances to query rewriting phase to be optimized before it is physically executed. The optimization is mostly eliminating unnecessary columns and operators, but the key optimization strategy of MapD is operator fusion that reduces intermediate output to avoid unnecessary materialization. For instance, MapD has a type of operator called \textit{RelCompound}, a combination of consecutive \textit{RelFilter} (selection), \textit{RelProject} (projection), and \textit{RelAggregate} (aggregation) to make the final output at once by coalescing them. Once the query rewriting is done, the plan is finally passed into physical query executor to generate the results and send back to the client.

\section{Selection Push-Down Decision Problem}\label{mapd_impl}
Although selection push-down is a simple yet powerful optimization technique, it generates intermediate result that should be materialized, which is directly written into memory and costs extra resources to process. For this reason, MapD does not perform any selection push-down to minimize the materialization which can cause performance degradation; the higher the cardinality of a pushed down selection is, the more data would be materialized and consequently slows down the system. Therefore, every selection predicates are just inside a single selection operator which is at the top of all join operators like the raw plan shown in top-left of figure \ref{fig:flowchart}. 

Yet, there still are cases that the selection push-down is effective in spite of the materialization when the size of intermediate output from the selection push-down is small. In this case, just a small amount of tuples will be materialized that would not cause significant overhead. Moreover, such a small intermediate output will exponentially lower the cost of join operations and results in performance boost. Our new query optimization strategy for MapD begins here: if the cardinality of a selection is too large, the query rewriter does not perform the selection push-down to avoid such a huge materialization. To determine whether to push a selection down or not, an accurate selectivity estimation method is required. This is where our novel selectivity estimation strategy comes in to help the system to choose better query execution plan which only contains helpful selections in the optimal plan. Plus, our approach is more suitable since MapD does not maintain any database statistics.

\begin{algorithm}[H]
  \caption{Selective Selection Push-Down}
  \label{alg:spd}
  \begin{algorithmic}[1]
    \Function{EvaluateAndPushDown}{$treeRA$}
    \State Initialize a priority queue $listRA$
    \For{\textbf{each} scan $R$ \textbf{in} $treeRA$}
    \If{$R.size > \texttt{PUSH\_DOWN\_MIN\_TABLE\_SIZE}$}
    \State $listRA.push(\{R, R.size\})$
    \EndIf
    \EndFor
    \State $listRA.pop()$ // remove the largest relation
    \For{\textbf{each} scan $R$ \textbf{in} $listRA$}
    \State $(conditions, columns) \gets $ {\small\textproc{ExtractPushDown}}($R$)
    \State $compound \gets {\small\textproc{CoalesceNodes}}(conditions, columns)$
    \State $outputOG \gets R.output$
    \State $outputOG.input \gets compound$
    \State $compound.input \gets R$
    \Try
    \State $maxSize \gets R.size * \texttt{PUSH\_DOWN\_MAX\_SELECTIVITY} $
    \State $resultSet \gets {\small\textproc{Execute}}(compound, true, maxSize)$
    \State ${\small\textproc{AddTemporaryTable}}(resultSet)$
    \State ${\small\textproc{UpdateTree}}(treeRA)$
    \Catch
    \State // revert selection push-down for $R$
    \State $outputOG.input = R$
    \EndTry
    \EndFor
    \EndFunction
    \Statex
    \Function{Execute}{$node$, $isSPD=false$, $maxSize=0$}
    \If{$ isSPD $}
    \State Create a temporary operator $\gamma_{\texttt{COUNT}}$
    \State $\gamma_{\texttt{COUNT}}.input \gets node$
    \State $count \gets {\small\textproc{Execute}}(\gamma_{\texttt{COUNT}}, false)$
    \If{$count > maxSize$}
    \State \textbf{throw} exception
    \EndIf
    \EndIf
    \State ... Perform physical query execution on $node$ ...
    \State \Return $resultSet$
    \EndFunction
  \end{algorithmic}
\end{algorithm}
\pagebreak
Algorithm \ref{alg:spd} shows an overview of our selection push-down extension for MapD. The function \textproc{EvaluateAndPushDown()} starts from the raw query execution plan $treeRA$ which has every predicates in a single selection operator at the top of all join operators. At the beginning of the function, it collects the size of each relation in a priority queue, ordered by the size of relation in descending order, if the size of a relation is greater than \texttt{PUSH\_DOWN\_MIN\_TABLE\_SIZE}; it is basically a fixed unsigned number which stands for the minimum number of tuples of a relation to avoid selection push-down when the size of a relation is already too small that push-down is not worthy. In case of TPC-H workbench dataset for example, we can avoid any selection push-down on tiny relations such as \textit{region} and \textit{nation} by setting this constant greater than their size. Once the size of relation $R$ is large enough, the information will be stored and sorted in the priority queue. In the end, we remove the first element in the queue that is the largest relation among them so that we do not push selection down for the relation to be probed.

The function \textproc{ExtractPushDown()} is then called to extract the selection predicates for each relation by looking up the selection operator at the top. Simultaneously, it also collects all the necessary columns from the selection and other projection operators above it. Projecting only necessary columns will significantly reduces the required resources while counting and executing the selection because MapD is a column-oriented database system. \textproc{CoalesceNodes()} is a predefined procedure in MapD that creates a fusion operator which is a combination of $\sigma_{conditions}(R)$ and $\pi_{columns}(R)$ to reduce the intermediate output. After this step, the query compiler inserts the fusion operator $compound$ into $treeRA$, right above the relation $R$.

When we pass $compound$ to the query executor, we also need to pass another parameter $maxSize$ derived from \texttt{PUSH\_DOWN\_MAX\_SELECTIVITY} which is a fixed ratio. $maxSize$ stands for the maximum selectivity of a pushed down selection that if the selectivity of a selection is too large, the selection will not be pushed down since it can cause performance degradation due to huge materialization. \texttt{PUSH\_DOWN\_MAX\_SELECTIVITY} could be defined as an unsigned number instead of a ratio depend on the size a relation so that we can directly pass that number instead calculated $maxSize$. We will discuss bit more about this later based on our experiments.

Inside the query executor, we insert a few lines of code to conditionally create an aggregate operator and connect it to $compound$ so that we can estimate the selectivity of the given selection on demand. We again call \textproc{Execute()} by passing $\gamma_{\textit{COUNT}}$ to compute the exact selectivity. Once it returns the selectivity, we compare the number with $maxSize$. Once the selectivity greater than the constant, we simply throw an exception to halt the current procedure immediately. Otherwise, the regular query execution will be performed that returns the actual result set of $compound$.

In case of exception, we simply reconnect the original output node of $R$ with $R$ so that we can discard the pushed down selection since it is determined as a worthless operation. On the other hand, we first store the result set as a temporary table so that it can be reused during the execution of the full query. Finally, we need to update the upper nodes affected by the selection push-down in $treeRA$. Those extracted predicates should now be removed from the selection operator at the top of all join operators, and schemas should be modified since we also pushed a projection down. We repeat the same process for each relation and in the end, only the helpful selections will be pushed down that consequently generate a better query execution plan.
\pagebreak
\chapter{Evaluation}\label{experiments}
\section{Setup}
The experiments are basically designed to compare the performance between with and without selection push-down with certain constraints. The experimental data is collected by running the same query with and without selection push-down several times and recorded the best of each. The performance between them is measured in terms of running time by using built-in timing option in MapD Core that returns the elapsed time from passing the query string to Apache Calcite to the end of its execution. We set \texttt{PUSH\_DOWN\_MIN\_TABLE\_SIZE} to a small yet large enough number to avoid selection push-down on two small relation, \textit{region} and \textit{nation}. 

For the experiments, MapD Core with selection push-down extension was set up in a single machine with two Intel(R) Xeon(R) CPU E5-2660 v4 @ 2.00GHz, eight DDR4 memory 32GB @ 2400 MHZ, and a single Tesla K80 GPU which has 24GB GDDR5 memory.

\section{Overheads from Selectivity Computation}
\begin{table*}[ht]
  \centering
  \caption{Overheads by size of relation (ms)}
  \label{table:overhead_sf}
  \begin{tabular}{|c|*{12}{c|}} \hline
    SF & \multicolumn{4}{c|}{50} & \multicolumn{4}{c|}{10} & \multicolumn{4}{c|}{1}\\ \hline
    Tables & O & PS & P & S & O & PS & P & S & O & PS & P & S\\ \hline
    GPU & 21 & 18 & 21 & 19 & 20 & 23 & 23 & 28 & 21 & 20 & 22 & 22\\ \hline
    CPU & 70 & 95 & 66 & 57 & 49 & 58 & 25 & 21 & 21 & 23 & 24 & 21\\ \hline
  \end{tabular}
\end{table*}

We initially measure overheads from selectivity computation which is executing an extra aggregate query to compute the exact selectivity to check whether this approach is really feasible or not. To do this, we first set \texttt{PUSH\_DOWN\_MAX\_SELECTIVITY} to 0 so that no selection is pushed down regardless of computed selectivity since MapD originally does not perform selection push-down in any case. This way, the final query execution plan will be just the same in both cases, but our approach will leave some overhead of running an extra query. We also perform the same procedure on both GPU and CPU to see the differences between them. We execute several queries on TPC-H benchmark dataset under specific circumstances.

We first run queries which join two tables, \textit{lineitem} and one of four other tables, \textit{orders}, \textit{partsupp}, \textit{part}, and \textit{supplier} with a highly selective predicate, in three different scale factors to see how our approach is affected by the size of relation. In case of \textit{orders}, for example, we run:
\begin{lstlisting}[caption={Overhead by size of relation}, label=sql:overhead_sf, basicstyle=\small\ttfamily]
SELECT COUNT(*)
FROM lineitem, orders
WHERE l_orderkey = o_orderkey
AND o_orderkey = 1;
\end{lstlisting}
We simply change the relation to be joined along with join condition for the other tables. Note that the size of each relation \textit{orders}, \textit{partsupp}, \textit{part}, \textit{supplier} is 1.5 million, 0.8 million, 0.2 million, and 10 thousand respectively when the scale factor is 1.

Table \ref{table:overhead_sf} shows the overhead for each case and the outputs are quite interesting. When running on CPU, the overhead obviously increases as the size of relation becomes larger. However, there is no such difference between outputs while running on GPU that the overhead is almost always consistent and rather smaller, less than 30 millisecond in any circumstance. 

We second run two queries over \textit{lineitem} and \textit{orders} with different selection predicates to change the actual selectivity and see whether it causes any significant overhead or not. To do this, we compare the running time of the query \ref{sql:overhead_sf} and its variant which has different selection predicate:
\begin{lstlisting}[caption={Overhead by selectivity}, label=sql:overhead_sel, basicstyle=\small\ttfamily]
SELECT COUNT(*)
FROM lineitem, orders
WHERE l_orderkey = o_orderkey
AND o_orderkey >= 1;
\end{lstlisting}
This will selects all tuples in \textit{orders} to maximize the gap in terms of selectivity. In the same manner, we set the scale factor to 50 that the selectivity difference between two queries becomes 75 million.

\begin{table}[h]
  \centering
  \caption{Overheads by selectivity (ms)}
  \label{table:overhead_sel}
  \begin{tabular}{|c|*{2}{c|}} \hline
    Selectivity & 1 & 75M\\ \hline
    GPU & 21 & 22\\ \hline
    CPU & 70 & 66\\ \hline
  \end{tabular}
\end{table}
However, there is no such difference, even while running on CPU as shown in Table \ref{table:overhead_sel}. Note that, when a typical database system executes a count function, it iterates through all the tuples and simply increase a counter whenever it finds a tuple which satisfies the given condition. Therefore, the only time consuming part is evaluating the selection predicates that the number and the type of attributes involved could be the factor of significant overhead for our strategy. This simple experiment shows that our approach actually does not produce any significant overhead other than running a simple aggregate query since the queries we run have only a single attribute with different operation, equality and inequality.

We throw another experiment to see whether the number of attributes actually affects the performance of our selectivity estimation strategy. Here is the query we run for this experiment:
\begin{lstlisting}[caption={Overhead by number of attributes}, label=sql:overhead_attr, basicstyle=\small\ttfamily]
SELCT COUNT(*)
FROM lineitem, orders
WHERE l_orderkey = o_orderkey
AND o_orderkey = 1
[AND o_custkey = 184500]
[AND o_orderstatus = 'O']
[AND o_totalprice = 218611.01];
\end{lstlisting}
We reuse the example query \ref{sql:overhead_sf} from the previous experiments, but this time we add additional selection predicate in bracket for each time. Note that the type of attribute in the third and the fourth predicate is \textit{TEXT} and \textit{DECIMAL(15,2)} respectively that may requires more time than \textit{INTEGER} does to evaluate.

\begin{table}[h]
  \centering
  \caption{Overheads by number of attributes (ms)}
  \label{table:overhead_attr}
  \begin{tabular}{|c|*{4}{c|}} \hline
    \# attributes & 1 & 2 & 3 & 4\\ \hline
    GPU & 21 & 23 & 23 & 18\\ \hline
    CPU & 70 & 91 & 273 & 339\\ \hline
  \end{tabular}
\end{table}
Table \ref{table:overhead_attr} shows the overheads by running queries with different number of attributes involved running on both CPU and GPU. While running on CPU, unfortunately the overhead increases as the number of attributes involved increases. The gap becomes even larger when evaluating the third and the fourth predicate as we expected. On the other hand, there is no such significant overhead while running on GPU that is even consistent like the previous experiments. This result is meaningful since our approach is originally designed to estimate the cardinality of a complicate selection.

These experiments show that a GPU-accelerated database system is more suitable for the novel selectivity estimation strategy since GPU is much better at running simple aggregate queries with its massively parallel architecture that does not cause any significant overhead, rather small and even consistent.

\section{Selection Push-Down Extension}
As shown in the previous chapter, the novel selectivity estimation strategy does not cause any significant overheads while running on GPU. Therefore, we set MapD Core run on GPU and see how selection push-down boosts the performance. To do this, we fix \texttt{PUSH\_DOWN\_MAX\_SELECTIVITY} to 1.0 to test all the different number of selectivity. 
Throughout all experiments, \textit{SPD} means that the selection push-down extension is enabled, i.e., selection is always pushed down since \texttt{PUSH\_DOWN\_MAX\_SELECTIVITY} is now set to 1.0, while \textit{no SPD} means that the extension is disabled so selection push-down is never performed. The queries ran were all simple count to minimize the number of columns to be loaded due to limited GPU memory in case of large scale factor because the more columns are projected, the more memory is required.

We first run the example query \ref{sql:overhead_sf} on the different scale factors of TPC-H benchmark dataset. This way, we can see whether selection push-down actually boosts the performance in all cases.

\begin{figure}[H]
  \centering
  \includegraphics[width=0.67\columnwidth]{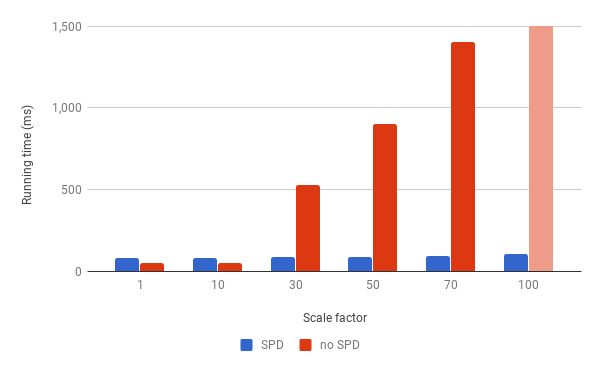}
  \caption{Running times by scale factor}
  \label{fig:spd_sf}
\end{figure}
Figure \ref{fig:spd_sf} shows the running times between with and without selection push-down for different scale factor. Even with such a small selectivity, there was no performance boost when the scale factor is small, 1 and 10. When size of relations are small, selection push-down is not effective due to data transfer limit between main memory and GPU and processing extra materialization. As of 30, however, selection push-down significantly reduces the running time, and the gap between with and without selection push-down becomes larger as the scale factor increases. Moreover, in case of scale factor of 100, it is not even able to execute the query without selection push-down on GPU. This is because it is not able to build full hash table for \textit{orders} on GPU memory. However, with push-down selection enabled, it dramatically reduces the number of tuples in \textit{orders} that we no longer need a huge hash table, and it takes only 0.1 seconds as a result. This shows that selection push-down becomes highly effective in larger dataset, especially with low selectivity which can guarantee performance boost.

From now on, we fix the scale factor to 50 that we can expect performance improvement at some point. The second experiment begins from measuring the running time with different selectivity by running this query:
\begin{lstlisting}[caption={Example to measure performance by selectivity on \textit{orders}}, label=sql:spd_sel_o, basicstyle=\small\ttfamily]
SELECT COUNT(*)
FROM lineitem, orders
WHERE l_orderkey = o_orderkey
AND o_orderkey <= $1;
\end{lstlisting}
We keep change the parameter \$1 so that we can find when the performance with selection push-down enabled overwhelms the one without it.

\begin{figure}[H]
  \centering
  \includegraphics[width=0.67\columnwidth]{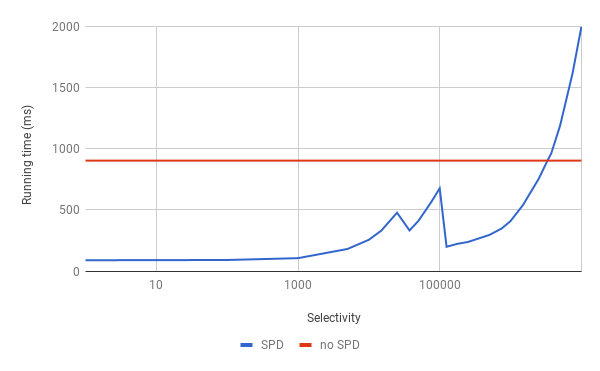}
  \caption{Running times by selectivity on \textit{orders}}
  \label{fig:spd_sel_o}
\end{figure}
Figure \ref{fig:spd_sel_o} shows the trends of running times for different selectivity on orders. Note that the running time without selection push-down is almost consistent since MapD originally executes selection once all join operations are done. On the other hand, the running time with selection push-down is exponential in log scale of selectivity, except the range between 5,000 and 100,000 which shows unexpected overheads. This happens because of changes in hash table layout from open addressing to perfect hashing inside MapD Core that open addressing utilizes memory better, less memory divergence for example, than perfect hashing at the limit between them. We ignore such behavior at this moment since selection push-down still brings performance improvement in that range, and hash table construction is different topic. Overall, selection push-down outperforms until the cardinality of pushed down selection on \textit{orders} becomes 3.3 million tuples, 4.4\% in ratio, which is the upper bound of where selection push-down is worth.

We next try another table \textit{partsupp} which is smaller than \textit{orders}. We find the upper bound first and then compare it with the one from \textit{orders} to figure out the correlation between the upper bound and the size of a relation by running a set of queries similar to the previous one:
\begin{lstlisting}[caption={Example to measure performance by selectivity on \textit{partsupp}}, label=sql:spd_sel_ps1, basicstyle=\small\ttfamily]
SELECT COUNT(*)
FROM lineitem, partsupp
WHERE l_partkey = ps_partkey
AND ps_partkey <= $1;
\end{lstlisting}

\begin{figure}[H]
  \centering
  \includegraphics[width=0.67\columnwidth]{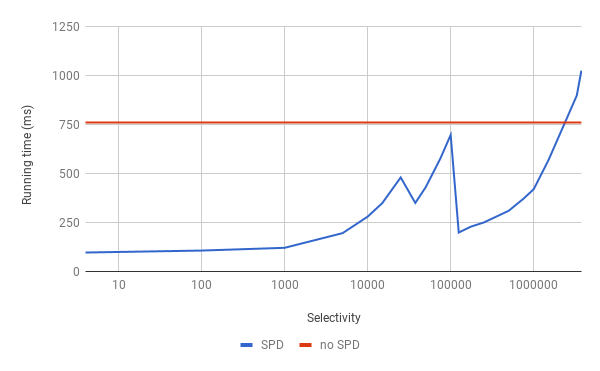}
  \caption{Running times by selectivity on \textit{partsupp}}
  \label{fig:spd_sel_ps1}
\end{figure}

While the trend shown in Figure \ref{fig:spd_sel_ps1} is very similar to figure \ref{fig:spd_sel_o}, the upper bound decreases to approximately 2.3 million tuples. This is because the room for the performance improvement by selection push-down is reduced as the size of relation decreases as we saw that selection push-down is not effective in case of a small scale factor. One more thing is that the upper bound of \textit{partsupp} is about 5.8\% in ratio that is actually greater than the one of \textit{orders}. This is the reason why we mentioned that \texttt{PUSH\_DOWN\_MAX\_SELECTIVITY} could be a fixed number and also a better heuristic instead of a ratio and passed directly to the query executor as the upper bound; as shown in the experiments so far, the actual upper bound of effective selection push-down decreases as the size of relation becomes smaller, but the ratio rather increases.

Before running another experiment over smaller relation, we join \textit{lineitem} and \textit{partsupp} over two predicates for this case by adding another join predicate to the previous query:
\begin{lstlisting}[caption={Example to measure performance by selectivity on \textit{partsupp} with two join predicates}, label=sql:spd_sel_ps1, basicstyle=\small\ttfamily]
SELECT COUNT(*)
FROM lineitem, partsupp
WHERE l_partkey = ps_partkey
AND l_suppkey = ps_suppkey
AND ps_partkey <= $1;
\end{lstlisting}
to see whether selection push-down is effective on complex join condition or not.

\begin{figure}[H]
  \centering
  \includegraphics[width=0.67\columnwidth]{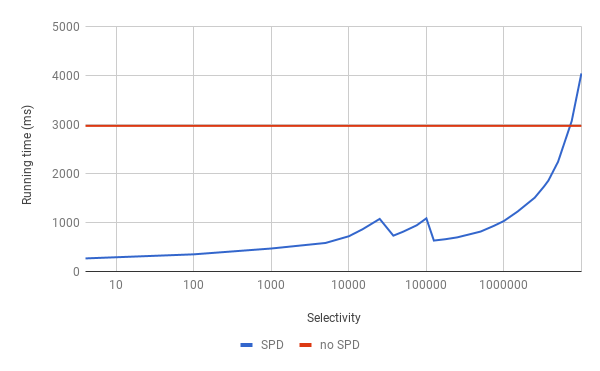}
  \caption{Running times by selectivity on \textit{partsupp} with two join predicates}
  \label{fig:spd_sel_ps2}
\end{figure}

With complex join between two tables, selection push-down becomes much more effective that the upper bound increases to approximately 7.5 million tuples, 18.75\% of \textit{partsupp}, as shown in figure \ref{fig:spd_sel_ps2}. This is because the hash table of \textit{partsupp} becomes much more complex which generally requires hashing twice. Hence, reducing the size of hash table with selection push-down is more effective as join predicates becomes complicate.

\begin{figure}[H]
  \centering
  \includegraphics[width=0.67\columnwidth]{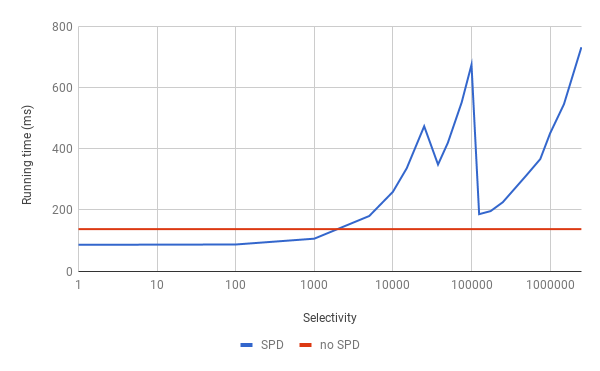}
  \caption{Running times by selectivity on \textit{part}}
  \label{fig:spd_sel_p}
\end{figure}
Figure \ref{fig:spd_sel_p} illustrates the trend of running times by different selectivity on \textit{part} which is even smaller than \textit{partsupp}. In this case, it is worth to perform selection push-down only when selectivity is less than a few thousand that is less than 0.1\% of the total number of tuples in this small relation. It is clear that selection push-down does not always guarantee performance boost and the less the size of relation is, the less effective selection push-down is.

We now throw another experiment to compare the performance by the number of attributes involved in a query. To to this, like what we did for measuring overheads of our strategy for different number of attributes involved, we also run the query \ref{sql:overhead_attr} by adding extra 5th predicate which shows interesting output:
\begin{lstlisting}[caption={Example to measure performance by number of attributes on \textit{orders}}, label=sql:spd_attr_o, basicstyle=\small\ttfamily]
SELCT COUNT(*)
FROM lineitem, orders
WHERE l_orderkey = o_orderkey
AND o_orderkey = 1
[AND o_custkey = 184500]
[AND o_orderstatus = 'O']
[AND o_totalprice = 218611.01]
[AND o_orderdate = DATE '1996-01-02'];
\end{lstlisting}

\begin{figure}[H]
  \centering
  \includegraphics[width=0.67\columnwidth]{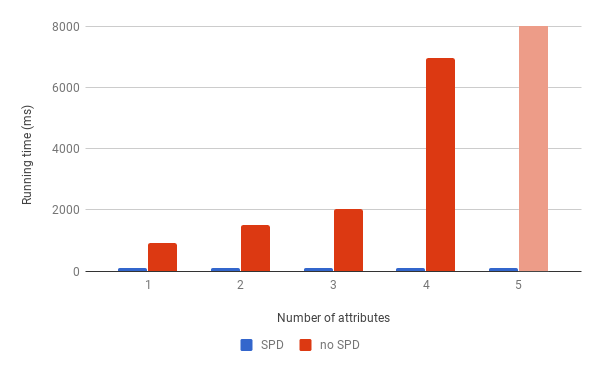}
  \caption{Running times by number of attributes on \textit{orders}}
  \label{fig:spd_attr_o}
\end{figure}
As shown in figure \ref{fig:spd_attr_o}, the running time without selection push-down becomes larger as the number of attributes involved increases, while the one with selection push-down is almost consistent as around 100 millisecond which is up to 68.2 times faster. Moreover, MapD Core even fails to build hash table on GPU memory once the number of attributes involved becomes 5 since the capacity of GPU memory is not large enough to hold all tuples. Selection push-down brings not only the performance improvement but also the more efficient memory usage when a selection has complicate predicates.

\begin{figure}[H]
  \centering
  \includegraphics[width=0.67\columnwidth]{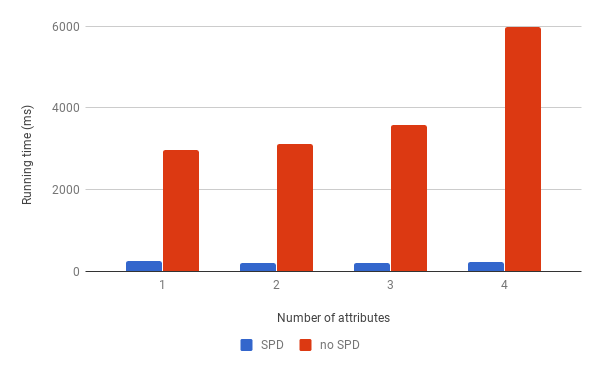}
  \caption{Running times by number of attributes on \textit{partsupp}}
  \label{fig:spd_attr_ps}
\end{figure}
Figure \ref{fig:spd_attr_ps} illustrates the trend of running times of the same type of experiment to perform selection push-down on \textit{partsupp} by running the query:
\begin{lstlisting}[caption={Example to measure performance by number of attributes on \textit{partsupp}}, label=sql:spd_attr_ps, basicstyle=\small\ttfamily]
SELCT COUNT(*)
FROM lineitem, partsupp
WHERE l_partkey = ps_partkey
AND l_suppkey = ps_suppkey
AND ps_partkey = 1
[AND ps_suppkey = 2]
[AND ps_availqty = 3325]
[AND ps_supplycost = 771.64]
\end{lstlisting}
Again, the gap between the running time with and without selection push-down increases as more attributes are involved in the query that the former is faster by up to 28.24 times. Although it is not noticeable in the figure, the running time with selection push-down even decreases slightly from 256 to 205 milliseconds while adding the second predicate since the actual selectivity decreases from 4 to 1. We can find that the number of attributes in selection predicates does not affect much the performance of selection push-down, but it is affected by the number of tuples selected instead.

Next experiment was designed to check whether the number of pushed down selections affects the performance or not. It is done by joining multiple tables consecutively, in the order of \textit{orders}, \textit{customer}, \textit{partsupp}, \textit{part} and \textit{supplier}, at each time with single selection predicate for each relation:
\begin{lstlisting}[caption={Example to measure performance by consecutively joining tables in different selectivity}, label=sql:spd_multi_ocpsps, basicstyle=\small\ttfamily]
SELECT COUNT(*)
FROM lineitem, orders
[, partsupp] [, part] [, customer] [, supplier]
WHERE l_orderkey = o_orderkey
[AND l_partkey = ps_partkey
 AND l_suppkey = ps_suppkey]
[AND ps_partkey = p_partkey]
[AND o_custkey = c_custkey]
[AND ps_suppkey = s_suppkey]
AND o_orderkey <= $1
[AND ps_partkey <= $2]
[AND p_partkey <= $3]
[AND c_custkey <= $4]
[AND s_suppkey <= $5];
\end{lstlisting}
Furthermore, we change the selectivity for each selection predicate like what we did with query \ref{sql:spd_sel_o} to check whether it introduces any extra overheads due to higher selectivity.

\begin{figure}[H]
  \centering
  \includegraphics[width=0.67\columnwidth]{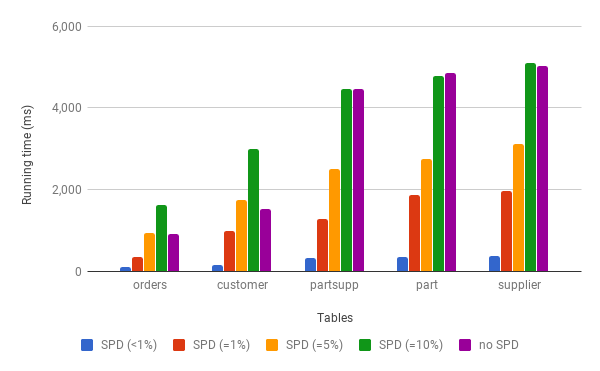}
  \caption{Running time by consecutively joining tables in different selectivity}
  \label{fig:spd_multi_ocpsps}
\end{figure}
The output is shown in Figure \ref{fig:spd_multi_ocpsps}. When the selectivity for each selection is small, it is stable and does not show any overheads and simply affected by the number of selected tuples from selection push-down. For instance, when the selectivity is 5\%, performance with selection push-down is worse than without selection push-down while joining \textit{orders} and \textit{customers}; joining \textit{lineitem} and \textit{orders} was better up to 4.4\% with selection push-down as previously shown in Figure \ref{fig:spd_sel_o}, and 5\% of customer would be worse since customer is far smaller than orders. When joining \textit{partsupp}, however, selection push-down is worth to do that joining \textit{partsupp} with selection push-down is almost twice faster than without selection push-down and effective up to 18.75\% as previously shown in Figure \ref{fig:spd_sel_ps2}. This shows that the number of pushed down selections does not introduce extra overhead and also proves that selection push-down can guarantee performance improvement below the certain upper bound.

\pagebreak
We finally run couple of TPC-H benchmark queries (cf. Appendix \ref{app:sql}). We chose three queries which include multiple tables and selection predicates so that we can expect any changes by enabling selection push-down extension.

\begin{figure}[H]
  \centering
  \includegraphics[width=0.67\columnwidth]{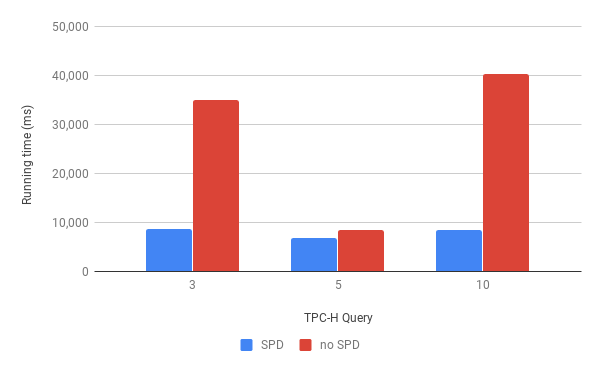}
  \caption{Performance of TPC-H benchmark queries (SF-50)}
  \label{fig:eg_tpch}
\end{figure}

As shown in figure \ref{fig:eg_tpch}, selection push-down boosts overall performance for chosen TPC-H benchmark queries up to 4.8$\times$. The improvement was not huge in case of query 5 due to relatively higher selectivity, 15.2\% of 75 million tuples, which involves huge materialization of intermediate data as a trade-off from selection push-down. 

Additionally, we run several SSB (Star Schema Benchmark) queries in scale factor of 80. Unlike TPC-H, the fact table \textit{lineorder} in SSB is huge that it takes more than 99\% of total dataset, which means the dimension tables are relatively smaller that selection push-down may not be effective. We run all the benchmark queries except the first cases that have restrictions on multiple dimensions.

\begin{figure}[H]
  \centering
  \includegraphics[width=0.67\columnwidth]{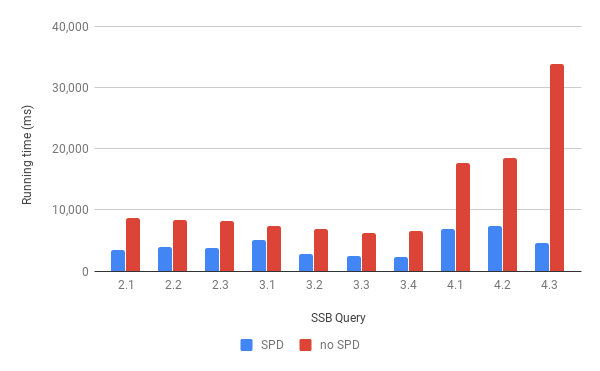}
  \caption{Performance of star schema benchmark queries (SF-80)}
  \label{fig:eg_ssb}
\end{figure}

Figure \ref{fig:eg_ssb} shows the performance of SSB queries. As the more tables and more selection predicates are involved, we get more performance improvement up to 7.3$\times$. The additional reason why the improvement in query 4.3 is larger than the one from the other queries is because it is a relatively more selective query. Selection push-down can also be effective when the fact table is huge even the dimension tables are small. Note that both SSB dataset with scale factor of 80 and TPC-H benchmark dataset with scale factor of 50 have almost 50 Gb dataset in total while the fact table \textit{lineorder} in SSB dataset which contains 780 million tuples is 2.6$\times$ larger than the largest table \textit{lineitem} in TPC-H benchmark dataset.
\pagebreak
\chapter{Related Work}\label{related_work}
Selectivity estimation is a critical part in query optimization that has been studied for decades but still remained as unsolved \cite{Chaudhuri:1998:OQO:275487.275492}. Histograms are the most frequently used type of statistics to provide better estimation of the selectivity \cite{Ioannidis:2003:HH:1315451.1315455}. While a histogram can provide an accurate estimation for a single attribute in a relation, it has difficulties with multiple attributes that the correlation between attributes is hard to figure out. Multi-dimensional histograms and linear algebra approaches are alternatives but becomes more problematic as the dimensionality increases \cite{Poosala:1997:SEW:645923.673638}. Sketches can be used for selectivity estimation problem \cite{Rusu:2007:SAS:1247480.1247503,Cormode:2005:IDS:1073713.1073718}, but also have difficulties on higher dimensional data as error exponentially increases. Samples are also widely used for cardinality estimation since such dimensionality is no longer a problem \cite{Wu:2012}. However, it is sensitive to skewed and sparse distribution of data due to its randomness of the sample generation, hence samples are more suitable to estimate trends and aggregations of data distribution \cite{Cormode:2012:SMD:2344400.2344401}. 

To overcome such selectivity estimation problems, we propose a novel selectivity estimation approach which computes the exact selectivity by executing an extra aggregate query. LEO, a LEarning Optimizer, \cite{Markl:2003:LAQ:1014767.1014781} is another type of query optimizer utilizes the exact cardinality of selections by monitoring query execution. While our approach computes the exact selectivity before executing the full query, LEO retrieves it during the query execution and adjusts database statistics to automatically correct the estimation error for the similar queries in the future. However, such deferred learning cannot effectively be applied to previously unanticipated queries that estimation could be wrong in this case as long as the estimates are based on database statistics and synopses. Additional approach is based on mid-query re-optimization \cite{Kabra:1998:EMR:276304.276315} that dynamically re-optimizes the query once the estimates are significantly different from the actual cardinalities to avoid executing sub-optimal plan. Changing a query execution plan however involves overhead to destroy obsoletes and build newly required structures such as hash tables while our approach builds only necessary structures from the beginning. Plus, the whole re-optimization process could be worthless if the chosen query execution plan is already an optimal.

Our strategy involves extra I/O access which is a trade-off for exact selectivity by running an aggregate query that can be a critical bottleneck in disk-based database systems. This issue can be resolved in modern database systems which uses main memory as a primary storage that significantly reduces I/O wait and are accelerated by GPU to boost the performance of the certain operations like aggregation and join with its highly parallelized architectures \cite{Bress:2014}. Nevertheless, data transfer between main memory and GPU can be performance bottleneck due to the limited bandwidth in PCIe bus that should be minimized \cite{He:2009:RQC:1620585.1620588}. MapD \cite{Mostak:2013} coalesces relational algebra operators so that it can combine multiple operations into a single pipeline to reduce data transfer and materialization and intermediate output. While selection push-down leaves materialized output in our evaluation, Funke et al. \cite{Funke:2018:PQP:3183713.3183734} proposed a pipelined query processing strategy which integrates multiple operations into a single kernel that does not leave any materialization. As long as the query compiler relies on database statistics or synopses, however, selectivity estimation can be erroneous that can lead to a sub-optimal query execution plan. Our strategy can outperform in such cases when the gap between an optimal and sub-optimal plan is greater than the cost of processing materialized data. Our approach can partially be applied to that pipelined query processing approach if we use only selectivity computation part; it guarantees accurate selectivity estimation and computation itself does not leave any materialization.
\chapter{Conclusions}\label{conclusions}
Selectivity estimation is an important part in query optimization yet an open question; estimates based on database statistics and synopses can be wrong especially in complex selection predicates that such wrong estimates let the query optimizer choose a sub-optimal plan and thus degrade the overall performance. In this paper, we propose a novel query optimization paradigm to solve selectivity estimation problems. Our approach is basically executing an extra aggregate query to count the cardinality of a selection by interacting with the query executor during the query optimization phase. This way, the exact selectivity can be achieved no matter how complex its predicate is that can result in an optimal query execution plan without any database statistics or synopses. We also suggest couple of optimization idea that considers materialization which is a trade-off for the benefit of selection push-down. We implement a selection push-down extension for a GPU-accelerated database systems, MapD, that can push only effective selections down based on simple heuristics with the computed selectivity for evaluation. We show that the overhead from selectivity computation is consistent and low in any circumstance while running on GPU, and performance improvement in certain cases especially in queries with complex selection predicates.
\pagebreak
% \nocite{*}
% \begin{footnotesize}
\bibliographystyle{elsarticle-num}
\bibliography{thesis}

%\bibliography{bib/general,bib/radprop,bib/pmradht,bib/flame,bib/pdf,bib/tao,bib/cmodes,bib/soot}	

% \end{footnotesize}

\appendix
\chapter{Example queries for TPC-H benchmark and star schema benchmark}\label{app:sql}
\begin{lstlisting}[caption={TPC-H Query 3}, basicstyle=\small\ttfamily]
SELECT l_orderkey, SUM(l_extendedprice * (1 - l_discount)) AS revenue, o_orderdate
FROM customer, orders, lineitem
WHERE c_mktsegment = 'BUILDING'
AND c_custkey = o_custkey
AND l_orderkey = o_orderkey
AND o_orderdate < DATE '1992-02-01' 
AND l_shipdate > DATE '1992-02-01' 
GROUP BY l_orderkey, o_orderdate 
ORDER BY revenue DESC, o_orderdate
LIMIT 10;
\end{lstlisting}

\begin{lstlisting}[caption={TPC-H Query 5}, basicstyle=\small\ttfamily]
SELECT n_name, SUM(l_extendedprice * (1 - l_discount)) AS revenue
FROM customer, orders, lineitem, supplier, nation, region
WHERE c_custkey = o_custkey
AND l_orderkey = o_orderkey
AND l_suppkey = s_suppkey
AND c_nationkey = s_nationkey
AND s_nationkey = n_nationkey
AND n_regionkey = r_regionkey
AND r_name = 'AFRICA'
AND o_orderdate >= DATE '1993-01-01'
AND o_orderdate < DATE '1993-01-01' + INTERVAL '1' YEAR
GROUP BY n_name
ORDER BY revenue DESC;
\end{lstlisting}

\begin{lstlisting}[caption={TPC-H Query 10}, basicstyle=\small\ttfamily]
SELECT c_custkey, c_name, SUM(l_extendedprice * (1 - l_discount)) AS revenue,
       c_acctbal, n_name, c_address, c_phone, c_comment
FROM customer, orders, lineitem, nation
WHERE c_custkey = o_custkey
AND l_orderkey = o_orderkey
AND o_orderdate >= DATE '1993-07-01'
AND o_orderdate < DATE '1993-07-01' + INTERVAL '3' MONTH
AND l_returnflag = 'R'
AND c_nationkey = n_nationkey
GROUP BY c_custkey, c_name, c_acctbal, c_phone, n_name, c_address, c_comment
ORDER BY revenue DESC
LIMIT 20;
\end{lstlisting}

\begin{lstlisting}[caption={SSB Query 2.1}, basicstyle=\small\ttfamily]
SELECT SUM(lo_revenue), d_year, p_brand1
FROM lineorder, part, supplier, ddate
WHERE lo_orderdate = d_datekey
AND lo_partkey = p_partkey
AND lo_suppkey = s_suppkey
AND p_category = 12
AND s_region = 1
GROUP BY d_year, p_brand1
ORDER BY d_year, p_brand1;
\end{lstlisting}

\begin{lstlisting}[caption={SSB Query 2.2}, basicstyle=\small\ttfamily]
SELECT SUM(lo_revenue), d_year, p_brand1
FROM lineorder, part, supplier, ddate
WHERE lo_orderdate = d_datekey
AND lo_partkey = p_partkey
AND lo_suppkey = s_suppkey
AND p_brand1 >= 2221
AND p_brand1 <= 2228
AND s_region = 2
GROUP BY d_year, p_brand1
ORDER BY d_year, p_brand1;
\end{lstlisting}

\begin{lstlisting}[caption={SSB Query 2.3}, basicstyle=\small\ttfamily]
SELECT SUM(lo_revenue), d_year, p_brand1
FROM lineorder, part, supplier, ddate
WHERE lo_orderdate = d_datekey
AND lo_partkey = p_partkey
AND lo_suppkey = s_suppkey
AND p_brand1 = 2239
AND s_region = 3
GROUP BY d_year, p_brand1
ORDER BY d_year, p_brand1;
\end{lstlisting}

\begin{lstlisting}[caption={SSB Query 3.1}, basicstyle=\small\ttfamily]
SELECT c_nation, s_nation, d_year, SUM(lo_revenue) AS revenue
FROM lineorder, customer, supplier, ddate
WHERE lo_custkey = c_custkey
AND lo_suppkey = s_suppkey
AND lo_orderdate = d_datekey
AND c_region = 2
AND s_region = 2
AND d_year >= 1992 AND d_year <= 1997
GROUP BY c_nation, s_nation, d_year
ORDER BY d_year ASC, revenue DESC;
\end{lstlisting}

\begin{lstlisting}[caption={SSB Query 3.2}, basicstyle=\small\ttfamily]
SELECT c_city, s_city, d_year, SUM(lo_revenue) AS revenue
FROM lineorder, customer, supplier, ddate
WHERE lo_custkey = c_custkey
AND lo_suppkey = s_suppkey
AND lo_orderdate = d_datekey
AND c_nation = 24
AND s_nation = 24
AND d_year >=1992 AND d_year <= 1997
GROUP BY c_city, s_city, d_year
ORDER BY d_year ASC, revenue DESC;
\end{lstlisting}

\begin{lstlisting}[caption={SSB Query 3.3}, basicstyle=\small\ttfamily]
SELECT c_city, s_city, d_year, SUM(lo_revenue) AS revenue
FROM lineorder, customer, supplier, ddate
WHERE lo_custkey = c_custkey
AND lo_suppkey = s_suppkey
AND lo_orderdate = d_datekey
AND (c_city = 231 OR c_city = 235)
AND (s_city = 231 OR s_city = 235)
AND d_year >=1992 AND d_year <= 1997
GROUP BY c_city, s_city, d_year
ORDER BY d_year ASC, revenue DESC;
\end{lstlisting}

\begin{lstlisting}[caption={SSB Query 3.4}, basicstyle=\small\ttfamily]
SELECT c_city, s_city, d_year, SUM(lo_revenue) AS revenue
FROM lineorder, customer, supplier, ddate
WHERE lo_custkey = c_custkey
AND lo_suppkey = s_suppkey
AND lo_orderdate = d_datekey
AND (c_city = 231 OR c_city = 235)
AND (s_city = 231 OR s_city = 235)
AND d_yearmonthnum = 199712
GROUP BY c_city,s_city,d_year
ORDER BY d_year ASC,revenue DESC;
\end{lstlisting}

\begin{lstlisting}[caption={SSB Query 4.1}, basicstyle=\small\ttfamily]
SELECT d_year, c_nation, SUM(lo_revenue-lo_supplycost) AS profit
FROM lineorder, supplier, customer, part, ddate
WHERE lo_custkey = c_custkey
AND lo_suppkey = s_suppkey
AND lo_partkey = p_partkey
AND lo_orderdate = d_datekey
AND c_region = 1
AND s_region = 1
AND (p_mfgr = 1 OR p_mfgr = 2)
GROUP BY d_year, c_nation
ORDER BY d_year, c_nation;
\end{lstlisting}

\begin{lstlisting}[caption={SSB Query 4.2}, basicstyle=\small\ttfamily]
SELECT d_year, s_nation, p_category, SUM(lo_revenue-lo_supplycost) AS profit
FROM lineorder, customer, supplier, part, ddate
WHERE lo_custkey = c_custkey
AND lo_suppkey = s_suppkey
AND lo_partkey = p_partkey
AND lo_orderdate = d_datekey
AND c_region = 1
AND s_region = 1
AND (d_year = 1997 OR d_year = 1998)
AND (p_mfgr = 1 OR p_mfgr = 2)
GROUP BY d_year, s_nation, p_category
ORDER BY d_year, s_nation, p_category;
\end{lstlisting}

\begin{lstlisting}[caption={SSB Query 4.3}, basicstyle=\small\ttfamily]
SELECT d_year, s_city, p_brand1, SUM(lo_revenue-lo_supplycost) AS profit
FROM lineorder, supplier, customer, part, ddate
WHERE lo_custkey = c_custkey
AND lo_suppkey = s_suppkey
AND lo_partkey = p_partkey
AND lo_orderdate = d_datekey
AND c_region = 1
AND s_nation = 24
AND (d_year = 1997 OR d_year = 1998)
AND p_category = 14
GROUP BY d_year, s_city, p_brand1
ORDER BY d_year, s_city, p_brand1;
\end{lstlisting}

% Ancillary material should be put in appendices, which appear after the
% bibliography. 

\end{document}